\documentclass[11pt]{article}
\usepackage[totalwidth=460truept,totalheight=600truept]{geometry}
\usepackage{latexsym,amssymb,amsmath,graphicx}
\usepackage[T1]{fontenc}

\def\theequation{\arabic{section}.\arabic{equation}}

\renewcommand{\theequation}{\thesection.\arabic{equation}}
\global\arraycolsep=1truept

\begin{document}

\hfill \hfill IFUP-TH 2007/18

\vskip 1.4truecm

\begin{center}
{\huge \textbf{Renormalization}}

{\huge \textbf{\large \vskip .1truecm}}

{\huge \textbf{Of Lorentz Violating Theories}}

\vskip 1.5truecm

\textsl{Damiano Anselmi and Milenko Halat}

\textit{Dipartimento di Fisica ``Enrico Fermi'', Universit\`{a} di Pisa, }

\textit{Largo Pontecorvo 3, I-56127 Pisa, Italy, }

\textit{and INFN, Sezione di Pisa, Pisa, Italy}

damiano.anselmi@df.unipi.it, milenko.halat@df.unipi.it
\end{center}

\vskip 2truecm

\begin{center}
\textbf{Abstract}
\end{center}

\bigskip

{\small We classify the unitary, renormalizable, Lorentz violating quantum
field theories of interacting scalars and fermions, obtained improving the
behavior of Feynman diagrams by means of higher space derivatives. Higher
time derivatives are not generated by renormalization. Renormalizability is
ensured by a ``weighted power counting'' criterion. The theories contain a
dimensionful parameter }$\Lambda _{L}${\small , yet a set of models are
classically invariant under a weighted scale transformation, which is
anomalous at the quantum level. Formulas for the weighted trace anomaly are
derived. The renormalization-group properties are studied.}

\vskip 1truecm

\vfill\eject

\section{Introduction}

\setcounter{equation}{0}

The set of power-counting renormalizable theories is considerably restricted
by the assumptions of unitarity, locality, causality and Lorentz invariance.
If we relax one or some of these assumptions we can enlarge the set of
renormalizable theories. However, usually the enlargement is too wide. For
example, there exist an infinite set of renormalizable nonunitary theories.
Improving the behavior of propagators at large momenta with the help of
higher-derivative kinetic terms \cite{slavnov} it is possible to define a
renormalizable higher-derivative version of every theory, including gravity 
\cite{stelle}. Relaxing locality can in principle make every theory
renormalizable, smoothing away the small distance singularities that
originate the UV divergences \cite{tomboulis}. Unitarity violations due to
higher derivatives can in some cases be traded for causality violations \cite
{mio,nostro}.

The purpose of this paper is to investigate the issue of renormalizability
in the presence of Lorentz violations, while preserving both locality and
unitarity. The UV behavior of propagators is improved with the help of
higher space derivatives. It is proved that, under certain conditions,
renormalization does not turn on terms with higher time derivatives, thus
preserving unitarity. Renormalizability follows from a modified
power-counting criterion, which weights time and space differently. The set
of consistent theories is still very restricted, yet considerably larger
than the set of Lorentz invariant theories. Renormalizable models exist in
arbitrary spacetime dimensions.

The quadratic terms that contain higher space derivatives, as well as
certain vertices, are multiplied by inverse powers of a scale $\Lambda _{L}$%
. Despite the presence of the dimensionful parameter $\Lambda _{L}$ certain
models have a weighted scale invariance, which is anomalous at the quantum
level. The weighted trace anomaly is worked out explicitly.

In this paper we concentrate on scalar and fermion theories, leaving the
study of gauge theories and gravity to separate publications. Lorentz
violating models with higher space derivatives might be useful to define the
ultraviolet limit of theories that are otherwise nonrenormalizable,
including quantum gravity, and allow to remove the divergences with a finite
number of independent couplings. Other domains where the models of this
paper might find applications are Lorentz violating extensions of the
Standard Model \cite{colladay}, effective field theory \cite{physica},
renormalization-group (RG) methods for the search of asymptotically safe
fixed points \cite{erg}, nonrelativistic quantum field theory for nuclear
physics \cite{wise1}, condensed matter physics and the theory of critical
phenomena \cite{pasquale}. Certain $\varphi ^{4}$-models that fall in our
class of renormalizable theories are useful to describe the critical
behavior at Lifshitz points \cite{lifshitz} and have been widely studied in
that context \cite{lifshitz2}, with a variety of applications to real
physical systems. Effects of Lorentz and CPT\ violations on stability and
microcausality have been studied \cite{lehnert}, as well as the induction of
Lorentz violations by the radiative corrections \cite{jackiw}. The
renormalization of gauge theories containing Lorentz violating terms has
been studied in \cite{gauge}. For a recent review on astrophysical
constraints on the Lorentz violation at high energy see ref. \cite{jacobson}.

The paper is organized as follows. In section 2 we study the
renormalizability of scalar theories, while in section 3 we include the
fermions. In section 4 we analyze the divergent parts of Feynman diagrams
and their subtractions. We prove the locality of counterterms and study the
renormalization algorithm to all orders. The one-loop divergences are
computed explicitly. In section 5 we analyze the renormalization structure
and the renormalization group. In section 6 we study the energy-momentum
tensor, the weighted scale invariance and the weighted trace anomaly. In
section 7 we generalize our results to nonrelativistic theories. Section 8
contains the conclusions. In the appendices we collect more observations
about the cancellation of subdivergences and the locality of counterterms,
and some expressions of Euclidean propagators in coordinate space.

\paragraph{Preliminaries.}

We use the dimensional-regularization technique whenever possible. Since the
analysis of divergences is the same in the Euclidean and Minkowskian
frameworks, we write our formulas directly in the Euclidean framework, which
is more explicit. Yet, with an abuse of language, we still speak of
``Lorentz symmetry'', since no confusion is expected to arise.

We first consider models where the $d$-dimensional spacetime manifold $M_{d}$
is split into the product $M_{\widehat{d}}\otimes M_{\overline{d}}$ of two
submanifolds, a $\widehat{d}$-dimensional submanifold $M_{\widehat{d}}$,
containing time and possibly some space coordinates, and a $\overline{d}$%
-dimensional space submanifold $M_{\overline{d}}$. Lorentz and rotational
symmetries in the two submanifolds are assumed. This kind of splitting could
be useful to describe specific physical situations (for example the presence
of a non isotropic medium in condensed matter physics), but here it is
mainly used as a starting point to illustrate our arguments in concrete
examples. Indeed, most Lorentz violating theories contain a huge number of
independent vertices, so it is convenient to begin with models where
unnecessary complicacies are reduced to a minimum. The extension of our
construction to the most general case, which is rather simple, will be
described later. In the same spirit, a number of discrete symmetries, such
as parity, time reversal, $\varphi\rightarrow - \varphi$, etc., are often
assumed.

To apply the dimensional-regularization technique, both submanifolds have to
be continued independently. The total continued spacetime manifold $M_{D}$
is therefore split into the product $M_{\widehat{D}}\otimes M_{\overline{D}}$%
, where $\widehat{D}=$ $\widehat{d}-\varepsilon _{1}$ and $\overline{D}=%
\overline{d}-\varepsilon _{2}$ are complex and $D=\widehat{D}+\overline{D}$.
Each momentum $p$ is split into ``first'' components $\widehat{p}$, which
live in $M_{\widehat{D}}$, and ``second'' components $\overline{p}$, which
live in $M_{\overline{D}}$: $p=(\widehat{p},\overline{p})$. The spacetime
index $\mu $ is split into hatted and barred indices: $\mu =(\widehat{\mu },%
\overline{\mu })$. Notations such as $\widehat{p}_{\widehat{\mu }}$, $%
\widehat{p}_{\mu }$ and $p_{\widehat{\mu }}$ refer to the same object, as
well as $\overline{p}_{\overline{\mu }}$, $\overline{p}_{\mu }$, $p_{%
\overline{\mu }}$. Frequently, Latin letters are used for the indices of the
barred components of momenta. Finally, $\overline{\triangle }\equiv 
\overline{\partial }_{i}\overline{\partial }_{i}$.

We say that $P_{k,n}(\widehat{p},\overline{p})$ is a weighted polynomial in $%
\widehat{p}$ and $\overline{p}$, of degree $k$ and weight $1/n$, where $k$
is a multiple of $1/n$, if $P_{k,n}(\xi ^{n}\widehat{p},\xi \overline{p})$
is a polynomial of degree $kn$ in $\xi $. Clearly, 
\[
P_{k_{1},n}(\widehat{p},\overline{p})P_{k_{2},n}(\widehat{p},\overline{p}%
)=P_{k_{1}+k_{2},n}(\widehat{p},\overline{p}). 
\]
We say that $H_{k,n}(\widehat{p},\overline{p})$ is a homogeneous weighted
polynomial in $\widehat{p}$ and $\overline{p}$, of degree $k$ and weight $%
1/n $, if $H_{k,n}(\lambda \widehat{p},\lambda ^{1/n}\overline{p})=\lambda
^{k}H_{k,n}(\widehat{p},\overline{p})$. It is straightforward to prove that a
weighted polynomial $P_{k,n}$ of degree $k$ can be expressed as a linear
combination of homogeneous weighted polynomials $H_{k^{\prime },n}$ of
degrees $k^{\prime }\leq k$.

\section{Renormalizability by weighted power counting}

\setcounter{equation}{0}

In this section we classify the renormalizable Lorentz violating scalar
field theories that can be constructed with the help of quadratic terms
containing higher space derivatives and prove that renormalization does not
generate higher time derivatives.

Consider a generic scalar field theory with a propagator defined by the
quadratic terms 
\begin{equation}
\mathcal{L}_{\mathrm{free}}=\frac{1}{2}(\widehat{\partial }\varphi )^{2}+%
\frac{1}{2\Lambda _{L}^{2n-2}}(\overline{\partial }^{n}\varphi )^{2},
\label{lkin}
\end{equation}
where $\Lambda _{L}$ is an energy scale. Up to total derivatives it is not
necessary to specify how the $2n$ derivatives $\overline{\partial }$
contract among themselves. The $n$ of (\ref{lkin}) should be understood as
the highest power of $\overline{\partial }$ that appears in the quadratic
terms of the total lagrangian. Other quadratic terms of the form 
\begin{equation}
\frac{a_{m}}{2\Lambda _{L}^{2m-2}}(\overline{\partial }^{m}\varphi
)^{2},\qquad m<n,  \label{am}
\end{equation}
could be present, or generated by renormalization. They are weighted
monomials of degrees $<2$ and weight $1/n$. For the purposes of
renormalization, it is convenient to consider such terms as ``interactions''
(two-leg vertices) and treat them perturbatively. Indeed, the counterterms
depend polynomially on the parameters $a_{m}$, because when the integral
associated with a graph is differentiated a sufficient number of times with
respect to the $a_{m}$'s it becomes overall convergent. The $a_{m}$%
-polynomiality of counterterms generalizes the usual polynomiality in the
masses. Thus we can assume that the propagator is defined by (\ref{lkin})
and treat every other term as a vertex. Then the propagator is the inverse
of a weighted homogeneous polynomial of degree $2$ and weight $1/n$. The
coefficient of the term $(\overline{\partial }^{n}\varphi )^{2}$ must be
positive, to have an action bounded from below in the Euclidean framework
or, equivalently, an energy bounded from below in the Minkowskian framework.

Label the vertices that have $N$ $\varphi $-legs with indices $\alpha $, to
distinguish different derivative structures. Each vertex of type $(N,\alpha)$
defines a monomial in the momenta of the fields. Denote the weighted degree
of such a monomial by $\delta _{N}^{(\alpha )}$. A vertex with $p_1$
derivatives $\widehat{\partial}$, $p_2$ derivatives $\overline{\partial}$
and $N$ $\varphi$-legs is symbolically written as 
\[
\left[\widehat{\partial}^{p_1}\overline{\partial}^{p_2}\varphi^N\right]_{%
\alpha} 
\]
and its weighted degree is 
\[
\delta _{N}^{(\alpha )}=p_1+\frac{p_2}{n}. 
\]

Consider a Feynman graph $G$ made of $L$ loops, $E$ external legs, $I$
internal legs and $v_{N}^{(\alpha )}$ vertices of type $(N,\alpha )$. The
integral associated with $G$ has the form 
\[
\mathcal{I}_{G}(k)=\int \frac{\mathrm{d}^{L\widehat{D}}\widehat{p}}{(2\pi
)^{L\widehat{D}}}\int \frac{\mathrm{d}^{L\overline{D}}\overline{p}}{(2\pi
)^{L\overline{D}}}\prod_{i=1}^{I}\mathcal{P}_{-2,n}^{(i)}(p,k)\prod_{j=1}^{V}%
\mathcal{V}_{\delta _{j},n}^{(j)}(p,k), 
\]
where $p$ are the loop momenta, $k$ are the external momenta, $\mathcal{P}%
_{-2,n}^{(i)}$ are the propagators, which have weighted degree $-2$, and $%
\mathcal{V}_{\delta _{j},n}^{(j)}$ are the vertices, with weighted degrees $%
\delta _{j}$. The integral measure $\mathrm{d}^{\widehat{D}}\widehat{p}\ 
\mathrm{d}^{\overline{D}}\overline{p}$ is a weighted measure of degree 
\textit{\DH }$\equiv \widehat{D}+\overline{D}/n$. Performing a rescaling $(%
\widehat{k},\overline{k})\rightarrow (\lambda \widehat{k},\lambda ^{1/n}%
\overline{k})$, accompanied by an analogous change of variables $(\widehat{p}%
,\overline{p})\rightarrow (\lambda \widehat{p},\lambda ^{1/n}\overline{p})$,
it is straightforward to prove that $\mathcal{I}_{G}(k)$ is a weighted function of
degree 
\[
L\hbox{\it \DH }-2I+\sum_{j=1}^{V}\delta _{j}=L\hbox{\it \DH }%
-2I+\sum_{(N,\alpha )}\delta _{N}^{(\alpha )}v_{N}^{(\alpha )}. 
\]
By the locality of counterterms, once the subdivergences of $G$ have been
inductively subtracted away, the overall divergent part of $G$ is a weighted
polynomial of degree 
\[
\omega (G)=L\hbox{\it \dj }-2I+\sum_{(N,\alpha )}\delta _{N}^{(\alpha
)}v_{N}^{(\alpha )} 
\]
in the external momenta, where \textit{\dj }$\equiv \widehat{d}+\overline{d}%
/n$. The usual relations 
\begin{equation}
L=I-V+1,\qquad E+2I=\sum_{(N,\alpha )}Nv_{N}^{(\alpha )},  \label{standard}
\end{equation}
allow us to write 
\begin{equation}
\omega (G)=d(E) +\sum_{(N,\alpha )}v_{N}^{(\alpha )}\left[ \delta
_{N}^{(\alpha )}-d(N)\right],
\end{equation}
where 
\begin{equation}
d(X)\equiv \hbox{\it \dj }\left( 1-\frac{X}{2}\right) +X\ ;  \label{nh}
\end{equation}

The theory is $i$) renormalizable, if it contains all vertices with $%
\delta_{N}^{(\alpha )}\leq d(N)$, and only those: $\omega (G)$ does not
increase when the number of vertices increases; $ii$) super-renormalizable,
if it contains all vertices with $\delta _{N}^{(\alpha )}<d(N)$, and only
those: $\omega (G)$ decreases when the number of vertices increases; $iii$)
strictly-renormalizable, if it contains all vertices with $\delta
_{N}^{(\alpha )}=d(N)$, and only those: $\omega (G)$ does not depend on $%
v_{N}^{(\alpha )}$; $iv$) nonrenormalizable, if it contains some vertices
with $\delta _{N}^{(\alpha )}>d(N)$: $\omega (G)$ increases when the number
of those vertices increases.

The vertices with $\delta _{N}^{(\alpha )}=d(N)$ are called ``weighted
marginal'', those with $\delta _{N}^{(\alpha )}<d(N)$ are called ``weighted
relevant'' and those with $\delta _{N}^{(\alpha )}>d(N)$ are called
``weighted irrelevant''.

By locality, $\delta _{N}^{(\alpha )}$ cannot be negative. Moreover,
polynomiality demands that there must exist a bound $N_{\mathrm{max}}$ on
the number of legs that the vertices can contain. It is easy to show that
these requirements are fulfilled if and only if 
\begin{equation}
\hbox{\it \dj }>2  \label{big}
\end{equation}
and the bound is 
\begin{equation}
N_{\mathrm{max}}=\left[ \frac{2\hbox{\it \dj }}{\hbox{\it \dj }-2}\right] ,
\label{nmax}
\end{equation}
where $[x]$ denotes the integral part of $x$. The existence of nontrivial
interactions ($N_{\mathrm{max}}\geq 3$) requires $\hbox{\it \dj }\leq 6$,
while the existence of nontrivial even interactions ($N_{\mathrm{max}}\geq
4 $) requires $\hbox{\it \dj }\leq 4$.

To complete the proof of renormalizability, observe that when $%
\delta_{N}^{(\alpha )}\leq d(N)$ the weighted degree of divergence $\omega
(G)$ of a graph $G$ satisfies 
\begin{equation}
\omega (G)\leq d(E).  \label{wg}
\end{equation}
The inequality (\ref{big}) ensures also that $\omega (G)$ decreases when the
number of external legs increases. Finally, since the vertices that subtract
the overall divergences of $G$ are of type $(E,\alpha)$ with $\delta
_{E}^{(\alpha )}=$ $\omega (G)$, it is straightforward to check that the
lagrangian contains all needed vertices. Indeed, (\ref{wg}) coincides with
the inequality satisfied by $\delta _{E}^{(\alpha )}$.

Now we prove that the renormalizable models just constructed are
perturbatively unitary, in particular that no higher time derivatives are
present, both in the kinetic part and in the vertices, and no higher time
derivatives are generated by renormalization. Indeed,\ a lagrangian term
with higher time derivatives would have $\delta _{N}^{(\alpha )}\geq 2$ for $%
N>2$ or $\delta _{2}^{(\alpha )}>2$ (terms with $N=1$ need not be
considered, since they cannot contain derivatives). This cannot happen in a
renormalizable theory, because (\ref{big}) and $\delta _{N}^{(\alpha )}\leq
d(N)$ imply $\delta _{N}^{(\alpha )}\leq 2$ in general and $\delta
_{N}^{(\alpha )}<2$ for $N>2$. In particular, true vertices ($N>2$) cannot
contain any $\widehat{\partial }$-derivative at all, because invariance
under the reduced Lorentz and rotational symmetries of $M_{\widehat{D}}$ and 
$M_{\overline{D}}$ exclude also terms containing an odd number of $\widehat{%
\partial }$'s or an odd number of $\overline{\partial }$'s. Similar
conclusions apply to the counterterms, because of (\ref{wg}). Therefore,
renormalization does not turn on higher time derivatives, as promised.

\paragraph{Weighted scale invariance.}

The strictly renormalizable models have the $\delta _{N}^{(\alpha )}=d(N)$.
Their lagrangian has the form 
\begin{equation}
\mathcal{L}_{(\widehat{d},\overline{d})}=\frac{1}{2}(\widehat{\partial }%
\varphi )^{2}+\frac{1}{2\Lambda _{L}^{2n-2}}(\overline{\partial }^{n}\varphi
)^{2}+\sum_{(N,\alpha )}\frac{\lambda _{(N,\alpha )}}{N!\Lambda
_{L}^{(n-1)\left( N+\widehat{d}-\widehat{d}N/2\right) }}\left[ {\overline{%
\partial }}^{nd(N)}{\varphi }^{N}\right] _{\alpha }.  \label{homo}
\end{equation}
Here $\left[ {\overline{\partial }}^{nd(N)}{\varphi }^{N}\right] _{\alpha }$
denotes a basis of lagrangian terms constructed with $N$ fields $\varphi $
and $nd(N)$ ${\overline{\partial }}$-derivatives acting on them, contracted
in all independent ways, and $\lambda _{(N,\alpha )}$ are dimensionless
couplings.

In the physical spacetime dimension $d=\widehat{d}+\overline{d}$ (the
continuation to complex dimensions will be discussed later) the classical
theories with lagrangians $\mathcal{L}_{(\widehat{d},\overline{d})}$ are
invariant under the weighted dilatation 
\begin{equation}
\hat{x}\rightarrow \hat{x}\ \mathrm{e}^{-\Omega },\qquad \bar{x}\rightarrow 
\bar{x}\ \mathrm{e}^{-\Omega /n},\qquad \varphi \rightarrow \varphi \ 
\mathrm{e}^{\Omega (\hbox{\it \dj }/2-1)},  \label{wd}
\end{equation}
where $\Omega$ is a constant parameter. Each lagrangian term scales with the
factor \textit{\dj }, compensated by the scaling factor of the integration
measure d$^{d}x$ of the action.

We call the models (\ref{homo}) homogeneous. Homogeneity is preserved by
renormalization, namely there exists a subtraction scheme in which no
lagrangian terms of weighted degrees smaller than $d(N)$ are turned on by
renormalization. This fact is evident using the dimensional-regularization
technique. Indeed, when $\delta _{N}^{(\alpha )}=d(N)$, the equality in (\ref
{wg}) holds, so $\omega (G)=d(E)=\delta _{E}^{(\alpha )}$.

The weighted scale invariance (\ref{wd}) is anomalous at the quantum level.
The weighted trace anomaly and its relation with the renormalization group
are studied in section 6.

Nonhomogeneous theories are those that contain both weighted marginal and
weighted relevant vertices. In these cases the weighted dilatation (\ref{wd}%
) is explicitly broken by the super-renormalizable vertices, and dynamically
broken by the anomaly.

Let us analyze some explicit examples, starting from the homogeneous models.

\paragraph{Homogeneous models.}

We begin with the $\varphi ^{4}$-theories. Setting $N_{\mathrm{max}}=4$ in (%
\ref{nmax}) we get 
\begin{equation}
\frac{10}{3}<\hbox{\it \dj }\leq 4.  \label{sola}
\end{equation}
One solution with \textit{\dj }$=4$ is the usual Lorentz-invariant $\varphi
^{4}$-theory in four dimensions ($\widehat{d}=\overline{d}=2$, $n=1$). A
simple Lorentz-violating solution is the model with $n=2$ described by the
lagrangian 
\begin{equation}
\mathcal{L}_{(2,4)}=\frac{1}{2}(\widehat{\partial }\varphi )^{2}+\frac{1}{%
2\Lambda _{L}^{2}}(\overline{\triangle }\varphi )^{2}+\frac{\lambda }{%
4!\Lambda _{L}^{2}}\varphi ^{4}.  \label{six}
\end{equation}
in six dimensions, with $\widehat{d}=2$, $\overline{d}=4$. The $\varphi^4$%
-theories with $n=2$ are used to describe the critical behavior at Lifshitz
points \cite{lifshitz,lifshitz2}.

It is clear that (\ref{sola}) admits infinitely many solutions for each
value of \textit{\dj }. For example, given a solution, such as (\ref{six}),
infinitely many others are obtained multiplying $\overline{d}$ and $n$ by a
common integer factor. For \textit{\dj }$=4$ we have the family of $2(n+1)$%
-dimensional theories 
\begin{equation}
\mathcal{L}_{(2,2n)}=\frac{1}{2}(\widehat{\partial }\varphi )^{2}+\frac{1}{%
2\Lambda _{L}^{2(n-1)}}(\overline{\partial }^{n}\varphi )^{2}+\frac{\lambda 
}{4!\Lambda _{L}^{2(n-1)}}\varphi ^{4}.  \label{22n}
\end{equation}
In general, for every Lorentz-invariant renormalizable theory there exists
an infinite family of Lorentz-violating renormalizable theories.

Let us now concentrate on four dimensions. The spacetime manifold can be
split as $(\widehat{d},\overline{d})=(0,4)$, $(1,3)$, $(2,2)$, $(3,1)$, $%
(4,0)$. There is no nontrivial solution with $\widehat{d}=0$. Indeed, (\ref
{nmax}) implies 
\[
N_{\mathrm{max}}=\left[ \frac{4}{2-n}\right] , 
\]
so $n$ can only be 1, which gives back the Lorentz invariant $\varphi ^{4}$%
-theory. For $\widehat{d}=1$ we get 
\[
N_{\mathrm{max}}=\left[ \frac{2\left( n+3\right) }{3-n}\right] . 
\]
The only nontrivial solution is $n=2$, which implies $N_{\mathrm{max}}=10$
and 
\begin{equation}
\mathcal{L}_{(1,3)}=\frac{1}{2}(\widehat{\partial }\varphi )^{2}+\frac{1}{%
2\Lambda _{L}^{2}}(\overline{\triangle }\varphi )^{2}+\frac{\lambda _{6}}{%
6!\Lambda _{L}^{4}}\varphi ^{4}(\overline{\partial }\varphi )^{2}+\frac{%
\lambda _{10}}{10!\Lambda _{L}^{6}}\varphi ^{10}.  \label{fi10}
\end{equation}
For $\widehat{d}=2$ we get $N_{\mathrm{max}}=2\left( n+1\right) $: every
integer $n>1$ defines a nontrivial solution in this case. The simplest
example is $(\widehat{d},\overline{d})=(2,2)$, $n=2$. Listing all allowed
vertices we get the theory 
\begin{equation}
\mathcal{L}_{(2,2)}=\frac{1}{2}(\widehat{\partial }\varphi )^{2}+\frac{1}{%
2\Lambda _{L}^{2}}\left( \overline{\triangle }\varphi \right) ^{2}+\frac{%
\lambda _{4}}{4!\Lambda _{L}^{2}}\varphi ^{2}(\overline{\partial }\varphi
)^{2}+\frac{\lambda _{6}}{6!\Lambda _{L}^{2}}\varphi ^{6}.  \label{4d}
\end{equation}
This model belongs to a family of \textit{\dj }$=3$, $(2+n)$-dimensional $%
\varphi ^{6}$-theories, whose lagrangian is 
\begin{equation}
\mathcal{L}_{(2,n)}=\frac{1}{2}(\widehat{\partial }\varphi )^{2}+\frac{1}{%
2\Lambda _{L}^{2(n-1)}}(\overline{\partial }^{n}\varphi )^{2}+\frac{\lambda
_{6}}{6!\Lambda _{L}^{2(n-1)}}\varphi ^{6},  \label{3d}
\end{equation}
when $n$ is odd and 
\begin{equation}
\mathcal{L}_{(2,n)}=\frac{1}{2}(\widehat{\partial }\varphi )^{2}+\frac{1}{%
2\Lambda _{L}^{2(n-1)}}(\overline{\partial }^{n}\varphi )^{2}+\frac{1}{%
4!\Lambda _{L}^{2(n-1)}}\sum_{\alpha }\lambda _{\alpha }\left[ {\overline{%
\partial }}^{n}{\varphi }^{4}\right] _{\alpha }+\frac{\lambda _{6}}{%
6!\Lambda _{L}^{2(n-1)}}\varphi ^{6},  \label{even}
\end{equation}
when $n$ is even. Observe that (\ref{3d}) includes the Lorentz-invariant $%
\varphi ^{6}$-theory in three spacetime dimensions, which is the case $n=1$.

For $\widehat{d}=3$ we get 
\[
N_{\mathrm{max}}=\left[ \frac{2\left( 3n+1\right) }{n+1}\right] . 
\]
The solution with $n=2$ has $N_{\mathrm{max}}=4$. However, this solution is
trivial, since its unique vertex would have just one $\overline{\partial }$%
-derivative. Instead, for every $n\geq 3$, $N_{\mathrm{max}}$ is equal to $5$%
. For example, the theory with $n=3$ is 
\[
\mathcal{L}_{(3,1)}=\frac{1}{2}(\widehat{\partial }\varphi )^{2}+\frac{1}{%
2\Lambda _{L}^{4}}\left( \overline{\partial }\overline{\triangle }\varphi
\right) ^{2}+\frac{\lambda _{3}^{\prime }}{3!\Lambda _{L}^{3}}\varphi ^{2}%
\overline{\triangle }^{2}\varphi +\frac{\lambda _{3}}{3!\Lambda _{L}^{3}}%
\varphi \left( \overline{\triangle }\varphi \right) ^{2}+\frac{\lambda _{4}}{%
4!\Lambda _{L}^{2}}\varphi ^{2}(\overline{\partial }\varphi )^{2}+\frac{%
\lambda _{5}}{5!\Lambda _{L}}\varphi ^{5}, 
\]
which is clearly unstable. Imposing the symmetry $\varphi \rightarrow
-\varphi $ we have the modified $\varphi ^{4}$-theory 
\[
\mathcal{L}_{(3,1)}^{\mathrm{even}}=\frac{1}{2}(\widehat{\partial }\varphi
)^{2}+\frac{1}{2\Lambda _{L}^{4}}\left( \overline{\partial }\overline{%
\triangle }\varphi \right) ^{2}+\frac{\lambda _{4}}{4!\Lambda _{L}^{2}}%
\varphi ^{2}(\overline{\partial }\varphi )^{2}, 
\]
which is stable for $\lambda _{4}>0$. Finally, for $\widehat{d}=4$ we get
again the Lorentz-invariant $\varphi ^{4}$-theory.

\paragraph{Nonhomogeneous models.}

Nonhomogeneous theories can be obtained from the homogeneous ones adding
all super-renormalizable terms, which are those that satisfy the strict
inequality $\delta _{N}^{(\alpha )}<d(N)$. For example, keeping the symmetry 
$\varphi \rightarrow -\varphi $, the nonhomogeneous extension of (\ref{six}%
) is just 
\[
\mathcal{L}_{(2,4)}^{\mathrm{nh}}=\frac{1}{2}(\widehat{\partial }\varphi
)^{2}+\frac{a}{2}(\overline{\partial }\varphi )^{2}+\frac{m^{2}}{2}\varphi
^{2}+\frac{1}{2\Lambda _{L}^{2}}(\overline{\triangle }\varphi )^{2}+\frac{%
\lambda }{4!\Lambda _{L}^{2}}\varphi ^{4}
\]
and the one of (\ref{4d}) is 
\[
\mathcal{L}_{(2,2)}^{\mathrm{nh}}=\frac{1}{2}(\widehat{\partial }\varphi
)^{2}+\frac{a}{2}(\overline{\partial }\varphi )^{2}+\frac{m^{2}}{2}\varphi
^{2}+\frac{1}{2\Lambda _{L}^{2}}\left( \overline{\triangle }\varphi \right)
^{2}+\frac{\lambda _{4}}{4!\Lambda _{L}^{2}}\varphi ^{2}(\overline{\partial }%
\varphi )^{2}+\frac{\lambda _{4}^{\prime }}{4!}\varphi ^{4}+\frac{\lambda
_{6}}{6!\Lambda _{L}^{2}}\varphi ^{6}.
\]

\paragraph{Splitting the spacetime manifold into the product of more
submanifolds.}

Instead of splitting the spacetime manifold into two submanifolds, we can
split it into the product of more submanifolds, eventually one for each
coordinate. This analysis covers the most general case. We still need to
distinguish a $\widehat{d}$-dimensional submanifold $M_{\widehat{d}}$
containing time from the $\overline{d}_{i}$-dimensional space submanifolds $%
M_{\overline{d}_{i}}$, $i=1,\ldots \ell $, so we write 
\[
M_{d}=M_{\widehat{d}}\otimes \prod_{i=1}^{\ell }M_{\overline{d}_{i}}. 
\]
Denote the space derivatives in the $i$th space subsector with $\overline{%
\partial }_{i}$ and assume that they have weights $1/n_{i}$. Then the
kinetic term of the lagrangian reads 
\[
\qquad \mathcal{L}_{\mathrm{kin}}=\frac{1}{2}(\widehat{\partial }\varphi
)^{2}+\frac{1}{2}\varphi P_{2}(\overline{\partial }_{i},\Lambda _{L})\varphi
, 
\]
where $P_{2}(\overline{\partial }_{i},\Lambda _{L})$ is the most general
weighted homogeneous polynomial of degree 2 in the spatial derivatives, $%
P_{2}(\lambda ^{1/n_{i}}\overline{\partial }_{i},\Lambda _{L})=\lambda
^{2}P_{2}(\overline{\partial }_{i},\Lambda _{L})$, invariant under rotations
in the subspaces $M_{\overline{d}_{i}}$. The $\Lambda _{L}$-dependence is
arranged so that $P_{2}$ has dimensionality 2. The previous analysis can be
repeated straightforwardly. It is easy to verify that the weighted
power-counting criterion works as before with 
\[
\hbox{\it \dj }=\widehat{d}+\sum_{i=1}^{\ell }\frac{\overline{d}_{i}}{n_{i}}%
. 
\]

\paragraph{Edge renormalizability.}

By edge renormalizable theories we mean theories where renormalization
preserves the derivative structure of the lagrangian, but the powers of the
fields are unrestricted. With scalars and fermions, such theories contain
arbitrary functions of the fields and therefore infinitely many independent
couplings. The notion of edge renormalizability is interesting in the
perspective to study gravity. Indeed, Einstein gravity is an example of
theory where all vertices have the same number of derivatives, but are
nonpolynomial in the fluctuation around flat space. Yet, diffeomorphism
invariance ensures that the number of invariants with a given dimensionality
in units of mass is finite. Therefore, in quantum gravity a polynomial
derivative structure is sufficient to reduce the arbitrariness to a finite
set of independent couplings.

Edge renormalizable theories are those where $\omega (G)$ does not decrease
when $E$ increases, rather it is independent of $E$. By formula (\ref{wg})
this means \textit{\dj }$=2$ ($N_{\mathrm{max}}=\infty $), in which case $%
\omega (G)$ is always equal to 2. Since \textit{\dj }$=2$, $\widehat{d}$ can
be either $0$ or $1$. The theories with $\widehat{d}=0$ contain higher time
derivatives, so they are not unitary. Thus we must take $\widehat{d}=1$. The
homogeneous theory in four dimensions has lagrangian 
\begin{equation}
\mathcal{L}=\mathcal{L}_{\mathrm{free}}+\mathcal{L}_{\mathrm{I}},
\label{lagrangian}
\end{equation}
where 
\[
\mathcal{L}_{\mathrm{free}}=\frac{1}{2}(\widehat{\partial }\varphi )^{2}+%
\frac{1}{2\Lambda _{L}^{4}}\left( \overline{\partial }\overline{\triangle }%
\varphi \right) ^{2} 
\]
and 
\begin{eqnarray}
\mathcal{L}_{\mathrm{I}}=V_{1}(\varphi )(\widehat{\partial }\varphi )^{2}+
&&V_{2}(\varphi )[(\partial _{i}\varphi )^{2}]^{3}+V_{3}(\varphi )\overline{%
\triangle }\varphi (\partial _{i}\varphi )^{2}(\partial _{j}\varphi
)^{2}+V_{4}(\varphi )(\partial _{i}\partial _{j}\varphi )(\partial
_{i}\partial _{j}\overline{\triangle }\varphi )  \nonumber \\
+V_{5}(\varphi )\overline{\triangle }^{2}\varphi (\partial _{i}\varphi
)^{2}+ &&V_{6}(\varphi )(\overline{\triangle }\varphi )^{3}+V_{7}(\varphi
)(\partial _{i}\overline{\triangle }\varphi )^{2}+V_{8}(\varphi )(\partial
_{i}\partial _{j}\partial _{k}\varphi )^{2}+V_{9}(\varphi )\overline{%
\triangle }^{3}\varphi ,  \label{plus}
\end{eqnarray}
where the $V_{i}$'s are unspecified functions of $\varphi $ with $%
V_{1}(\varphi )=\mathcal{O}(\varphi )$, $V_{4}(\varphi ),V_{7}(\varphi
),V_{8}(\varphi ),V_{9}(\varphi )=\mathcal{O}(\varphi ^{2})$.

The lagrangian of the most general nonhomogeneous theory is (\ref
{lagrangian}) with 
\[
\mathcal{L}_{\mathrm{free}}=\frac{1}{2}(\widehat{\partial }\varphi )^{2}-%
\frac{1}{2}\varphi \left( a\overline{\triangle }+b\frac{\overline{\triangle }%
^{2}}{\Lambda _{L}^{2}}+\frac{\overline{\triangle }^{3}}{\Lambda _{L}^{4}}%
\right) \varphi 
\]
and $\mathcal{L}_{\mathrm{I}}$ equal to (\ref{plus}) plus 
\[
V_{10}(\varphi )+V_{11}(\varphi )\overline{\triangle }\varphi
+V_{12}(\varphi )\overline{\triangle }^{2}\varphi +V_{13}(\varphi )(%
\overline{\triangle }\varphi )^{2}+V_{14}(\varphi )[(\partial _{i}\varphi
)^{2}]^{2}, 
\]
with $V_{11}(\varphi ),V_{12}(\varphi )=\mathcal{O}(\varphi ^{2})$, $%
V_{13}(\varphi )=\mathcal{O}(\varphi )$.

\section{Inclusion of fermions}

\setcounter{equation}{0}

In this section we classify the models of interacting fermions and scalars.
We start from pure fermionic theories, with quadratic lagrangian 
\[
\mathcal{L}_{\text{free}}=\overline{\psi }\widehat{\partial }\!\!\!\slash%
\,\psi +\frac{1}{\Lambda _{L}^{n-1}}\overline{\psi }{\overline{\partial }%
\!\!\!\slash\,}^{n}\psi , 
\]
where $n$ is the maximal number of ${\overline{\partial }}$-derivatives. The
propagator 
\[
\frac{-i\widehat{p}\!\!\!\slash \,+(-i)^{n}\frac{\overline{p}\!\!\!\slash %
\,^{n}}{\Lambda _{L}^{n-1}}}{\widehat{p}^{2}+\frac{(\overline{p}^{2})^{n}}{%
\Lambda _{L}^{2n-2}}}, 
\]
is, in momentum space, a weighted function of degree $-$1. The loop-integral
measure is, as usual, a weighted measure of degree \textit{\dj }. For the
purposes of renormalization, the kinetic terms with fewer than $n$ $%
\overline{\partial }$-derivatives can be treated as vertices.

Label the vertices that have $2N$ $\psi $-$\overline{\psi }\,$-legs by means
of indices $\alpha $ and denote their weighted degree with $\delta
_{N}^{(\alpha )}$. Consider a diagram $G$ with $2E$ external $\psi $-$%
\overline{\psi }\,$-legs, constructed with $v_{N}^{(\alpha )}$ vertices of
type $(N,\alpha )$. Once the subdivergences have been subtracted away, its
overall divergence is a weighted polynomial of degree 
\[
\omega (G)=\hbox{\it \dj }-E(\hbox{\it \dj }-1)+\sum_{(N,\alpha
)}v_{N}^{(\alpha )}\left[ \delta _{N}^{(\alpha )}-\hbox{\it \dj }\left(
1-N\right) -N\right] 
\]
in the external momenta. Renormalizability demands 
\begin{equation}
\delta _{N}^{(\alpha )}\leq \hbox{\it \dj }\left( 1-N\right) +N\equiv
d_{F}(N).  \label{ft}
\end{equation}
Polynomiality demands 
\[
\hbox{\it \dj }>1, 
\]
in which case the maximal number of external $\psi $-$\overline{\psi }\,$%
-legs is 
\[
N_{\max }=\left[ \frac{\hbox{\it \dj }}{\hbox{\it \dj }-1}\right] . 
\]
Pure fermionic homogeneous models have strictly renormalizable vertices,
namely those with $\delta _{N}^{(\alpha )}=d_{F}(N)$. Their lagrangian has
the form 
\[
\mathcal{L}=\overline{\psi }\widehat{\partial }\!\!\!\slash\,\psi +\frac{1}{%
\Lambda _{L}^{n-1}}\overline{\psi }{\overline{\partial }\!\!\!\slash\,}%
^{n}\psi +\sum_{(N,\alpha )}\frac{\lambda _{(N,\alpha )}}{(N!)^{2}\Lambda
_{L}^{(n-1)(N-\widehat{d}-N\widehat{d})}}\left[ {\overline{\partial }}%
^{nd_{F}(N)}\overline{\psi }^{N}\,\psi ^{N}\right] _{\alpha }. 
\]
Here $\left[ {\overline{\partial }}^{nd_{F}(N)}\overline{\psi }^{N}\,\psi
^{N}\right] _{\alpha }$ denotes a basis of lagrangian terms constructed with 
$N$ fields $\psi $, $N$ fields $\overline{\psi }$ and $nd_{F}(N)$ $\overline{%
\partial }$-derivatives, invariant under the reduced Lorentz symmetry. For
simplicity, we can assume also invariance under parities in both portions of
spacetime.

Let us concentrate on four spacetime dimensions. The Lorentz split $(1,3)$
gives $N_{\max }=1+[n/3]$, which admits infinitely many nontrivial
solutions, beginning from $n=3$. For example, the $n=3$ and $n=6$ theories
read 
\begin{eqnarray*}
\mathcal{L}_{(1,3)} &=&\overline{\psi }\widehat{\partial }\!\!\!\slash\psi +%
\frac{1}{\Lambda _{L}^{2}}\overline{\psi }\overline{\Delta }\overline{%
\partial }\!\!\!\slash\,\psi +\sum_{\alpha }\frac{\lambda _{\alpha }}{%
\Lambda _{L}^{2}}\left[ \overline{\psi }^{2}\,\psi ^{2}\right] _{\alpha }, \\
\mathcal{L}_{(1,3)}^{\prime } &=&\overline{\psi }\widehat{\partial }\!\!\!%
\slash\psi +\frac{1}{\Lambda _{L}^{5}}\overline{\psi }~\overline{\Delta }%
^{3}\psi +\sum_{\alpha }\frac{\lambda _{\alpha }}{\Lambda _{L}^{5}}\left[ {%
\overline{\partial }}^{3}\overline{\psi }^{2}\,\psi ^{2}\right] _{\alpha
}+\sum_{\alpha }\frac{\lambda _{\alpha }^{\prime }}{\Lambda _{L}^{5}}\left[ 
\overline{\psi }^{3}\psi ^{3}\right] _{\alpha },
\end{eqnarray*}
respectively. The Lorentz splits $(2,2)$ and $(3,1)$ do not admit
nontrivial solutions, since $N_{\max }=1$ in those cases.

Now we study the models containing coupled scalars and fermions. It is
important to note that when different types of fields are involved, they
must have the same $n$. We classify the vertices with labels $(N_{\psi
},N_{\varphi },\alpha )$, where $2N_{\psi }$ is the number of $\psi $-$%
\overline{\psi }\,$-legs, $N_{\varphi }$ is the number of $\varphi $-legs
and $\alpha $ is an extra label that distinguishes vertices with different
structures. Call $\delta _{(N_{\psi },N_{\varphi })}^{(\alpha )}$ the
weighted degree of the $\alpha $-th vertex. Consider a diagram $G$ with $%
2E_{\psi }$ external $\psi $-$\overline{\psi }\,$-legs, $E_{\varphi }$
external $\varphi $-legs and $v_{(N_{\psi },N_{\varphi })}^{(\alpha )}$
vertices of type $(N_{\psi },N_{\varphi },\alpha )$. Once the subdivergences
have been subtracted away, the overall divergent part of $G$ a is a weighted
polynomial of degree 
\begin{eqnarray*}
\omega (G) &=&\hbox{\it \dj }-E_{\psi }(\hbox{\it \dj }-1)-\frac{E_{\varphi }%
}{2}(\hbox{\it \dj }-2) \\
&&+\sum_{(N_{\psi },N_{\varphi },\alpha )}v_{(N_{\psi },N_{\varphi
})}^{(\alpha )}\left[ \delta _{(N_{\psi },N_{\varphi })}^{(\alpha )}-%
\hbox{\it \dj }\left( 1-N_{\psi }-\frac{N_{\varphi }}{2}\right) -N_{\psi
}-N_{\varphi }\right] .
\end{eqnarray*}
in the external momenta. Renormalizability demands 
\[
\delta _{(N_{\psi },N_{\varphi })}^{(\alpha )}\leq \hbox{\it \dj }\left(
1-N_{\psi }-\frac{N_{\varphi }}{2}\right) +N_{\psi }+N_{\varphi }\equiv
d(N_{\psi },N_{\varphi }).
\]
Because $\delta _{(N_{\psi },N_{\varphi })}^{(\alpha )}$ is nonnegative,
the numbers of fermionic and bosonic legs are bound by the inequality 
\[
N_{\psi }(\hbox{\it \dj }-1)+\frac{N_{\varphi }}{2}(\hbox{\it \dj }-2)\leq %
\hbox{\it \dj }.
\]
Polynomiality demands \textit{\dj }$>2$.

The homogeneous models have a lagrangian of the form 
\begin{eqnarray*}
\mathcal{L} &=&\overline{\psi }\widehat{\partial }\!\!\!\slash\psi +\frac{%
\eta}{\Lambda _{L}^{n-1}}\overline{\psi }\overline{\partial }\!\!\!\slash%
^{n}\psi +\frac{1}{2}(\widehat{\partial }\varphi )^{2}+\frac{1}{2\Lambda
_{L}^{2n-2}}(\overline{\partial }^{n}\varphi )^{2} \\
&&+\sum_{_{(N_{\psi },N_{\varphi },\alpha )}}\frac{\lambda _{(N_{\psi
},N_{\varphi },\alpha )}}{N_{\varphi }!(N_{\psi }!)^{2}\Lambda
_{L}^{(n-1)(N_{\varphi }+N_{\psi }+\widehat{d}-\widehat{d}N_{\psi }-\widehat{%
d}N_{\varphi }/2)}}\left[ {\overline{\partial }}^{nd(N_{\psi },N_{\varphi })}%
\overline{\psi }^{N_{\psi }}\psi ^{N_{\psi }}\varphi ^{N_{\varphi }}\right]
_{\alpha }.
\end{eqnarray*}
In four dimensions the splitting $(1,3)$ has a unique nontrivial solution,
which is the model (\ref{fi10}) coupled to fermions. It has $n=2$ and its
lagrangian reads 
\begin{eqnarray*}
\mathcal{L}_{(1,3)} &=&\overline{\psi }\widehat{\partial }\!\!\!\slash\psi +%
\frac{\eta}{\Lambda _{L}}\overline{\psi }\overline{\Delta }\psi +\frac{1}{2}(%
\widehat{\partial }\varphi )^{2}+\frac{1}{2\Lambda _{L}^{2}}(\overline{%
\Delta }\varphi )^{2}+\frac{\lambda _{2}}{2\Lambda _{L}^{2}}\varphi ^{2}(%
\overline{\psi }\overleftrightarrow{\overline{\partial }\!\!\!\slash}\psi )+%
\frac{\lambda _{2}^{\prime }}{2\Lambda _{L}^{2}}\varphi ^{2}\overline{%
\partial }\cdot (\overline{\psi }\overline{\gamma }\psi ) \\
&&+\frac{\lambda _{4}}{4!\Lambda _{L}^{3}}\varphi ^{4}\overline{\psi }\psi +%
\frac{\lambda _{6}}{6!\Lambda _{L}^{4}}\varphi ^{4}(\overline{\partial }%
\varphi )^{2}+\frac{\lambda _{10}}{10!\Lambda _{L}^{6}}\varphi ^{10}.
\end{eqnarray*}

The splitting $(2,2)$ admits infinitely many solutions. The simplest one is
the theory with $n=2$, symmetric under $\varphi \leftrightarrow -\varphi $,
that couples (\ref{4d}) to fermions: 
\[
\mathcal{L}_{(2,2)}=\overline{\psi }\widehat{\partial }\!\!\!\slash\,\psi +%
\frac{\eta}{\Lambda _{L}}\overline{\psi }\overline{\Delta }\psi +\frac{1}{2}(%
\widehat{\partial }\varphi )^{2}+\frac{1}{2\Lambda _{L}^{2}}(\overline{%
\Delta }\varphi )^{2}+\frac{\lambda _{2}}{2\Lambda _{L}}\varphi ^{2}%
\overline{\psi }\psi +\frac{\lambda _{4}}{4!\Lambda _{L}^{2}}\varphi ^{2}(%
\overline{\partial }\varphi )^{2}+\frac{\lambda _{6}}{6!\Lambda _{L}^{2}}%
\varphi ^{6}, 
\]
The splitting $(3,1)$ admits, again, infinitely many solutions.

\section{Renormalization}

\setcounter{equation}{0}

In this section we study the structure of Feynman diagrams, their
divergences and subdivergences, and the locality of counterterms. For
definiteness, we work with scalar fields, but the conclusions are general.

\paragraph{One-loop.}

Consider the most general one-loop Feynman diagram $G$, with $E$ external
legs, $I$ internal legs and $v_{N}^{(\alpha )}$ vertices of type $(N,\alpha
) $ and weighted degree $\delta _{N}^{(\alpha )}$. Collectively denote the
external momenta by $k$. The divergent part of $G$ can be calculated
expanding the integral in powers of $k$. We obtain a linear combination of
contributions of the form 
\begin{equation}
\mathcal{I}_{\mu _{1}\cdots \mu _{2r}|j_{1}\cdots j_{2s}}^{(I,n)}\widehat{k}%
_{\nu _{1}}\cdots \widehat{k}_{\nu _{u}}\ \overline{k}_{i_{1}}\cdots 
\overline{k}_{i_{v}},  \label{comba}
\end{equation}
where 
\[
\mathcal{I}_{\mu _{1}\cdots \mu _{2r}|j_{1}\cdots j_{2s}}^{(I,n)}=\int \frac{%
\mathrm{d}^{\widehat{D}}\widehat{p}}{(2\pi )^{\widehat{D}}}\int \frac{%
\mathrm{d}^{\overline{D}}\overline{p}}{(2\pi )^{\overline{D}}}\frac{\widehat{%
p}_{\mu _{1}}\cdots \widehat{p}_{\mu _{2r}}\ \overline{p}_{j_{1}}\cdots 
\overline{p}_{j_{2s}}}{\left( \widehat{p}^{2}+\left( \overline{p}^{2}\right)
^{n}/\Lambda _{L}^{2(n-1)}+m^{2}\right) ^{I}}. 
\]
To avoid infrared problems we insert a mass $m$ in the denominators. For the
purposes of renormalization, it is not necessary to think of $m$ as the real
mass. It can be considered as a fictitious parameter, introduced to
calculate the divergent part of the integral and set to zero afterwards. The
real mass, as well as the other parameters $a_{m}$ of (\ref{am}), can be
treated perturbatively, so they are included in the set of ``vertices''.

From the weighted power-counting analysis of section 2 we know that the
numerator of (\ref{comba}), namely 
\[
\widehat{p}_{\mu _{1}}\cdots \widehat{p}_{\mu _{2r}}\ \overline{p}%
_{j_{1}}\cdots \overline{p}_{j_{2s}}\ \widehat{k}_{\nu _{1}}\cdots \widehat{k%
}_{\nu _{u}}\ \overline{k}_{i_{1}}\cdots \overline{k}_{i_{v}}, 
\]
is a weighted monomial $P_{q,n}(\widehat{p},\widehat{k};\overline{p},%
\overline{k})$ of weight $1/n$ and degree 
\[
q=u+2r+\frac{v}{n}+\frac{2s}{n}=\sum_{(N,\alpha )}\delta _{N}^{(\alpha
)}v_{N}^{(\alpha )}. 
\]
At one loop the number of vertices equals the number of propagators. Using (%
\ref{standard}) and $\delta _{N}^{(\alpha )}\leq d(N)$ we get 
\begin{equation}
u+\frac{v}{n}\leq 2\left( I-r-\frac{s}{n}\right) +E\left( 1-\frac{%
\hbox {\it
\dj }}{2}\right) .  \label{condition}
\end{equation}
By symmetric integration, we can write 
\[
\mathcal{I}_{\mu _{1}\cdots \mu _{2r}|j_{1}\cdots j_{2s}}^{(I,n)}=\delta
_{\mu _{1}\cdots \mu _{2r}}^{(1)}\delta _{j_{1}\cdots j_{2s}}^{(2)}\mathcal{I%
}_{r,s}^{(I,n)},\quad \mathcal{I}_{r,s}^{(I,n)}=\int \frac{\mathrm{d}^{%
\widehat{D}}\widehat{p}}{(2\pi )^{\widehat{D}}}\int \frac{\mathrm{d}^{%
\overline{D}}\overline{p}}{(2\pi )^{\overline{D}}}\frac{\left( \widehat{p}%
^{2}\right) ^{r}\left( \overline{p}^{2}\right) ^{s}}{\left( \widehat{p}%
^{2}+\left( \overline{p}^{2}\right) ^{n}/\Lambda _{L}^{2(n-1)}+m^{2}\right)
^{I}}, 
\]
where $\delta _{\mu _{1}\cdots \mu _{2r}}^{(1)}$ and $\delta _{j_{1}\cdots
j_{2s}}^{(2)}$ are appropriately normalized completely symmetric tensors
constructed with the Kronecker tensors of $M^{\widehat{D}}$ and $M^{%
\overline{D}}$, respectively. Performing the change of variables 
\begin{equation}
\overline{p}_{i}=\overline{p}^{\prime }{}_{i}\left( \frac{\Lambda _{L}^{2}}{%
\overline{p}^{\prime 2}}\right) ^{(n-1)/(2n)},  \label{transformation}
\end{equation}
the integral $\mathcal{I}_{r,s}^{(I,n)}$ can be calculated using the
standard formulas of the dimensional-regularization technique. We obtain 
\begin{eqnarray*}
\mathcal{I}_{r,s}^{(I,n)} &=&\frac{1}{n}\Lambda _{L}^{(2s+\overline{D}%
)(n-1)/n}\int \frac{\mathrm{d}^{\widehat{D}}\widehat{p}}{(2\pi )^{\widehat{D}%
}}\int \frac{\mathrm{d}^{\overline{D}}\overline{p}^{\prime }}{(2\pi )^{%
\overline{D}}}\frac{(\widehat{p}^{2})^{r}(\overline{p}^{\prime 2})^{(2s+%
\overline{D}-n\overline{D})/(2n)}}{(\widehat{p}^{2}+\overline{p}^{\prime
2}+m^{2})^{I}} \\
&=&\frac{\Lambda _{L}^{(2s+\overline{D})(n-1)/n}(m^{2})^{r-I+s/n+%
\hbox {\it
\DJ }/2}\Gamma \left( \frac{2s+\overline{D}}{2n}\right) \Gamma \left( \frac{%
2r+\widehat{D}}{2}\right) \Gamma \left( I-r-\frac{s}{n}-\frac{\hbox{\it \DJ }%
}{2}\right) }{n(4\pi )^{D/2}\Gamma (\widehat{D}/2)\Gamma \left( \overline{D}%
/2\right) \Gamma \left( I\right) }.
\end{eqnarray*}
The factor $1/n$ is due to the Jacobian determinant of the transformation (%
\ref{transformation}). The singularities occur for 
\begin{equation}
I\leq r+\frac{s}{n}+\frac{\hbox{\it \dj }}{2}.  \label{divcond}
\end{equation}
Combining this inequality with (\ref{condition}) we find that the divergent
contributions satisfy 
\begin{equation}
u+\frac{v}{n}\leq \hbox{\it \dj }+E\left( 1-\frac{\hbox{\it \dj }}{2}\right)
=d(E).  \label{max}
\end{equation}
The counterterms are a $P_{u+v/n,n}(\widehat{k},\overline{k})$: 
\[
\frac{1}{\varepsilon }\widehat{k}_{\nu _{1}}\cdots \widehat{k}_{\nu _{u}}\ 
\overline{k}_{i_{1}}\cdots \overline{k}_{i_{v}},\qquad \mathrm{where}\text{ }%
\varepsilon =\hbox{\it \dj }-\hbox{\it \DJ }=\varepsilon _{1}+\frac{%
\varepsilon _{2}}{n}. 
\]
Thus (\ref{max}) ensures that the divergent terms can be subtracted away
renormalizing the fields and couplings of the initial lagrangian. Observe
that while the poles are proportional to $1/\varepsilon $, the residues of
the poles can depend on $\varepsilon _{1}$ and $\varepsilon _{2}$
separately. We know that taking a sufficient number of derivatives with
respect to the masses, the external momenta and the parameters $a_{m}$ of (%
\ref{am}), the integral becomes convergent. Therefore, the finite parts are
regular in the limits $\varepsilon _{1},\varepsilon _{2}\rightarrow 0$,
which can be safely taken in any preferred order. Objects such as $%
\varepsilon _{1}/\varepsilon $ and $\varepsilon _{2}/\varepsilon $ multiply
only local terms, so they parametrize different scheme choices and never
enter the physical quantities. These observations generalize immediately to
all orders. We define the minimal subtraction schemes as the schemes where 
\[
\varepsilon _{1}=\alpha \varepsilon ,\qquad \varepsilon _{2}=n(1-\alpha
)\varepsilon , 
\]
with $\alpha $=constant, and only the pure poles in $\varepsilon $ are
subtracted away, with no finite contributions.

\paragraph{Overall divergences and subdivergences.}

Before considering Lorentz violating theories to all orders in the loop
expansion it is convenient to briefly review the usual classification of
divergences and the proof of locality of counterterms \cite{collins} in
Lorentz symmetric theories. Consider the $L$-loop integral 
\[
\mathcal{I}(k)=\int \prod_{i=1}^{L}\frac{\mathrm{d}^{D}p^{(i)}}{(2\pi )^{D}}%
Q(p^{(1)},\ldots ,p^{(L)};k) 
\]
with Lorentz invariant propagators $1/(p^{2}+m^{2})$, where $k$ denotes the
external momenta. The ultraviolet behavior of $\mathcal{I}(k)$ is studied
letting any (sub)set of the momenta $p^{(1)},\ldots ,p^{(L)}$ tend to
infinity with the same velocity. Proper subsets of the momenta test the
presence of subdivergences, while the whole set tests the presence of
overall divergences. $i$) When any subconvergence fails, counterterms
corresponding to the divergent subdiagrams have to be included to subtract
the subdivergences. $ii$) Once all subdivergences are removed, the
subtracted integral $\mathcal{I}_{\mathrm{sub}}(k)$ can still be overall
divergent. Taking an appropriate number $M$ of derivatives with respect to
the external momenta $k$ the integral $\partial _{k}^{M}\mathcal{I}_{\mathrm{%
sub}}(k)$ becomes overall convergent. This proves the locality of
counterterms.

The overlapping divergences can be tested sending momenta to infinity with
different velocities. For example, rescale $p_{1},\ldots ,p_{L}$ as $\lambda
p_{1},\ldots ,\lambda p_{l},\lambda ^{2}p_{l+1},\ldots ,\lambda ^{2}p_{L}$.
This test, however, is already covered by the previous ones, since there is
always a (sub)set $s_{\mathrm{fast}}$ of momenta tending to infinity with
maximal velocity. In the example just given, $s_{\mathrm{fast}%
}=(p_{l+1},\ldots ,p_{L})$. The other momenta $s_{\mathrm{slow}}$ grow
slower, so they can be considered fixed in the first analysis and taken to
infinity at a second stage. Weinberg's theorem \cite{weinberg} ensures that
when $s_{\mathrm{fast}}$ tends to infinity the behavior of the relevant
subintegral is governed by power counting and can generate logarithmic
corrections depending on the momenta of $s_{\mathrm{slow}}$. Then, when $s_{%
\mathrm{slow}}$ tends to infinity the behavior of the integral over $s_{%
\mathrm{slow}}$ is still governed by power counting, because the corrections
due to the integrals over $s_{\mathrm{fast}}$ do not affect the powers of
the momenta $s_{\mathrm{slow}}$. Thus the power-counting analysis done in
steps $i$) and $ii$) suffices.

Now we generalize the analysis to Lorentz violating theories. We say that
the components $\widehat{p}$ and $\overline{p}$ of each momentum are
rescaled with the same ``weighted velocity'' when 
\[
\widehat{p}\rightarrow \lambda \widehat{p},\qquad \overline{p}\rightarrow
\lambda ^{1/n}\overline{p}. 
\]
Step $i$) is modified studying the convergence when any subset of momenta
tend to infinity with the same weighted velocity. Whenever a subconvergence
fails the counterterms associated with the divergent subdiagrams have to be
included. Once the subdivergences are subtracted away, step $ii$) consists
of taking an appropriate number of ``weighted derivatives'' (see below) with
respect to the external momenta, to eliminate the overall divergences. It is
easy to check that this procedure automatically takes care of the
overlapping divergences.

\paragraph{Weighted Taylor expansion.}

Every Taylor expansion

\[
f(\widehat{k},\overline{k})=\sum_{u=0}^{\infty }\sum_{v=0}^{\infty }\frac{%
f_{\nu _{1}\cdots \nu _{u},i_{1}\cdots i_{v}}}{u!v!}\widehat{k}_{\nu
_{1}}\cdots \widehat{k}_{\nu _{u}}\ \overline{k}{}_{i_{1}}\cdots \overline{k%
}{}_{i_{v}} 
\]
can be rearranged into a ``weighted Taylor expansion'' 
\[
f(\widehat{k},\overline{k})=\sum_{\ell =0}^{\infty }\frac{1}{\ell !}f^{(\ell
)}(\widehat{k},\overline{k}), 
\]
where 
\[
f^{(\ell )}(\widehat{k},\overline{k})=\sum_{u=0}^{[\ell /n]}\frac{\ell !}{%
u!(\ell -nu)!}f_{\nu _{1}\cdots \nu _{u},i_{1}\cdots i_{\ell -nu}}\widehat{k}%
_{\nu _{1}}\cdots \widehat{k}_{\nu _{u}}\ \overline{k}{}_{i_{1}}\cdots 
\overline{k}{}_{i_{\ell -nu}} 
\]
is a weighted homogeneous polynomial of degree $\ell /n$: 
\[
f^{(\ell )}(\lambda \widehat{k},\lambda ^{1/n}\overline{k})=\lambda ^{\ell
/n}f^{(\ell )}(\widehat{k},\overline{k}). 
\]
The $\ell $-th weighted derivatives with weight $1/n$ are the coefficients $%
f_{\nu _{1}\cdots \nu _{u},i_{1}\cdots i_{\ell -nu}}$.

The weighted Taylor expansion is useful to subtract the overall divergences.
The overall-subtracted version of an integral whose weighted degree of
divergence is $\omega $ reads 
\[
\int \frac{\mathrm{d}^{L\widehat{D}}\widehat{p}}{(2\pi )^{L\widehat{D}}}%
\frac{\mathrm{d}^{L\overline{D}}\overline{p}}{(2\pi )^{L\overline{D}}}\left[
Q(\widehat{p},\overline{p};\widehat{k},\overline{k})-\sum_{\ell =0}^{n\omega
}\frac{1}{\ell !}Q^{(\ell )}(\widehat{p},\overline{p};\widehat{k},\overline{k%
})\right] , 
\]
where $Q^{(\ell )}$ denotes the $\ell $-th homogeneous polynomial of the
weighted Taylor expansion of $Q$ in $\widehat{k},\overline{k}$.

\paragraph{Subtraction algorithm.}

Consider an $L$-loop diagram with $V$ vertices and $I$ propagators. The
integrand, which we denote with $Q_{G}$, is a ratio of weighted polynomials
and has degree equal to $d_{Q}\equiv \sum_{(N,\alpha )}\delta _{N}^{(\alpha
)}v_{N}^{(\alpha )}-2I$. The integral $\mathcal{I}$ is a weighted function
of degree $d_{\mathcal{I}}=d_{Q}+$\textit{\DH }$L$. It has the form 
\begin{equation}
\mathcal{I}=\int \frac{\mathrm{d}^{L\widehat{D}}\widehat{p}}{(2\pi )^{L%
\widehat{D}}}\int \frac{\mathrm{d}^{L\overline{D}}\overline{p}}{(2\pi )^{L%
\overline{D}}}Q_{G}(\widehat{p},\overline{p},k),  \label{integral}
\end{equation}
where $\widehat{p}$ and $\overline{p}$ collectively denote the components of
the momenta circulating in the loops, while $k=(\widehat{k},\overline{k})$
collectively denotes the external momenta. The overall degree of divergence
of $\mathcal{I}$ is $\omega (G)=d_{Q}+$\textit{\dj }$L$.

The subtraction of divergences can be arranged according to the following
table: 
\begin{equation}
\begin{tabular}{ccc}
$Q_{G}(\widehat{p},\overline{p};\widehat{k},\overline{k})$ & \phantom{CCC} & 
$-\sum_{\gamma \in \Gamma }\overline{Q}_{\gamma }(\widehat{p},\overline{p};%
\widehat{k},\overline{k})$ \\ 
&  &  \\ 
$-\sum_{\ell =0}^{n\omega (G)}\frac{1}{\ell !}Q_{G}^{(\ell )}(\widehat{p},%
\overline{p};\widehat{k},\overline{k})$ & \phantom{CCC} & $\sum_{\ell
=0}^{n\omega (G)}\frac{1}{\ell !}\sum_{\gamma \in \Gamma }\overline{Q}%
_{\gamma }^{(\ell )}(\widehat{p},\overline{p};\widehat{k},\overline{k})$%
\end{tabular}
\label{table}
\end{equation}
Here $\Gamma $ denotes the set of divergent subdiagrams $\gamma $ of the
diagram $G$. The rational function $\overline{Q}_{\gamma }$ is obtained
replacing the subintegrand with the appropriate, truncated, weighted Taylor
expansion in the external momenta of $\gamma $. In the arrangement of (\ref
{table}) subdivergences are subtracted row-wise. Overall divergences are
subtracted column-wise.

A potential caveat comes from certain ``extra subdivergences'',those that
occur when a subdiagram $\gamma ^{\prime }$ is convergent in $Q_{G}$, but
becomes divergent in one of the $\overline{Q}_{\gamma }$'s. Then $\gamma
^{\prime }$ does not belong to $\Gamma $, so its subdivergence is not
subtracted row-wise. Nevertheless, it is easy to show that the extra
subdivergences are automatically subtracted column-wise in (\ref{table}).
Details and an explicit example are given in appendix A.

Thus, once the subdivergences have been subtracted away, the divergent part
of every Feynman diagram is a weighted polynomial of degree $\omega (G)$
(second row of (\ref{table})) and can be removed renormalizing the
lagrangian (\ref{lagrangian}).

\section{Renormalization structure and renormalization group}

\setcounter{equation}{0}

In this section we study the renormalization group. We illustrate it first
in the \textit{\dj }$=4$ models (\ref{22n}). For the reasons that we explain
below, it is convenient to parametrize the bare lagrangian as 
\begin{equation}
\mathcal{L}_{(2,2n)\mathrm{B}}=\frac{1}{2}(\widehat{\partial }\varphi _{%
\mathrm{B}})^{2}+\frac{1}{2\Lambda _{L\mathrm{B}}^{2(n-1)}}(\overline{%
\partial }^{n}\varphi _{\mathrm{B}})^{2}+\frac{\lambda _{\mathrm{B}}}{%
4!\Lambda _{L\mathrm{B}}^{(n-1)(2-\varepsilon _{2}/n)}}\varphi _{\mathrm{B}%
}^{4}  \label{bare}
\end{equation}
with 
\begin{equation}
\varphi _{\mathrm{B}}=Z_{\varphi }^{1/2}\varphi ,\qquad \Lambda _{L\mathrm{B}%
}=Z_{\Lambda }\Lambda _{L},\qquad \lambda _{\mathrm{B}}=\lambda \mu
^{\varepsilon }Z_{\lambda },\qquad \varepsilon \equiv \varepsilon _{1}+\frac{%
\varepsilon _{2}}{n}.  \label{reno}
\end{equation}
Observe that \textit{\DJ }$=4-\varepsilon $. The weighted scale invariance (%
\ref{wd}) can be extended to a transformation that rescales also $\mu $: 
\begin{equation}
\hat{x}\rightarrow \hat{x}\ \mathrm{e}^{-\Omega },\qquad \bar{x}\rightarrow 
\bar{x}\ \mathrm{e}^{-\Omega /n},\qquad \varphi \rightarrow \varphi \ 
\mathrm{e}^{\Omega (\hbox{\it \DJ }-2)/2},\qquad \mu \rightarrow \mu \mathrm{%
e}^{\Omega }.  \label{wd2}
\end{equation}
The invariance under this transformation is not a symmetry. It just tells us
that at the quantum level the weighted scale invariance (\ref{wd}) is
equivalent to a $\mu $-rescaling. What is important in (\ref{wd}) and (\ref
{wd2}) is that $\Lambda _{L}$ is unmodified. Because of (\ref{wd2}), every
renormalization constant in (\ref{reno}) is just a function of $\lambda $
(otherwise it could also depend on evanescent powers of the ratio $\mu
/\Lambda _{L}$). Thus, in the minimal subtraction scheme the $\lambda $-beta
function has the usual form 
\[
\mu \frac{\mathrm{d}\lambda }{\mathrm{d}\mu }=\widehat{\beta }_{\lambda
}=-\varepsilon \lambda +\beta (\lambda ). 
\]
The finiteness of $\widehat{\beta }_{\lambda }$ proves that all poles
contained in $Z_{\lambda }$ are inverse powers of $\varepsilon $.

In more detail, let us consider the contribution of a graph $G$ with $E$
external legs, $I$ propagators and $V$ vertices to the generating functional
of one-particle irreducible diagrams. Such a contribution has the schematic
form 
\[
\mathcal{I}=\int \mathrm{d}^{D}x\frac{\lambda ^{V}\mu ^{V\varepsilon }}{%
\Lambda _{L}^{V(n-1)(2-\varepsilon _{2}/n)}}G\varphi ^{E}, 
\]
where $G$ denotes the value of the Green function. The dimensionality of $G$%
\ in units of mass is 
\[
\lbrack G]=D\left( V-\frac{E}{2}+1\right) +E-4V, 
\]
while its weighted degree is 
\[
\omega (G)=[G]-\delta [G]=4-E+\Delta \omega (G), 
\]
where 
\[
\Delta \omega (G)=-\varepsilon \left( V-\frac{E}{2}+1\right) ,\qquad \delta
[G]=\left( 2-\frac{\varepsilon _{2}}{n}\right) (n-1)\left( V-\frac{E}{2}%
+1\right) . 
\]
Recalling that $\mathcal{I}$ is invariant under the weighted scale
transformation (\ref{wd2}), we find that $G$ transforms as 
\begin{equation}
G\rightarrow \mathrm{e}^{\Omega \omega (G)}G.  \label{ag}
\end{equation}
Once the subdivergences have been inductively subtracted away, the divergent
part $G_{\mathrm{div}}$ is a weighted polynomial of degree $4-E$ in the
external momenta. Matching the dimensionality and the weighted rescaling (%
\ref{ag}) we find 
\[
G_{\mathrm{div}}=P_{4-E,n}(\widehat{\partial },\overline{\partial };\Lambda
_{L})\Lambda _{L}^{\delta [G]}\mu ^{\Delta \omega (G)}, 
\]
where $P_{4-E,n}(\widehat{\partial },\overline{\partial };\Lambda _{L})$ is
a homogeneous weighted polynomial of degree $4-E$ and dimensionality equal
to its degree. The corresponding lagrangian counterterm reads 
\[
\mathcal{I}_{\mathrm{div}}=-\int \mathrm{d}^{D}x\left( \frac{\lambda \mu
^{\varepsilon }}{\Lambda _{L}^{(n-1)(2-\varepsilon _{2}/n)}}\right) ^{V}\mu
^{\Delta \omega (G)}\Lambda _{L}^{\delta [G]}[P_{4-E,n}(\widehat{\partial },%
\overline{\partial };\Lambda _{L})]\varphi ^{E}, 
\]
where $[P]$ means that the derivatives contained in $P$ act on the scalar
legs $\varphi ^{E}$ as appropriate. In particular, summing up all
contributions for $E=4$, we get 
\[
-\int \mathrm{d}^{D}x\frac{\lambda \mu ^{\varepsilon }}{\Lambda
_{L}^{(n-1)(2-\varepsilon _{2}/n)}}\varphi ^{4}\sum_{L=1}^{\infty
}c_{L}\lambda ^{L}, 
\]
where $c_{L}$ are divergent constants. Thus the renormalization constant of $%
\lambda $ is a power series in $\lambda $, 
\[
Z_{\lambda }=1-\sum_{L=1}^{\infty }c_{L}\lambda ^{L}, 
\]
with no spurious dependence on $\mu /\Lambda _{L}$. The same conclusion
holds for the other renormalization constants. We have 
\[
\mu \frac{\mathrm{d}\Lambda _{L}}{\mathrm{d}\mu }=\eta _{L}\Lambda
_{L},\qquad \eta _{L}(\lambda )=-\frac{\mathrm{d}\ln Z_{\Lambda }}{\mathrm{d}%
\ln \mu }. 
\]

The Callan-Symanzik equation has the same form as usual. Calling 
\[
G_{k}(\widehat{x}_{1},\cdots ,\widehat{x}_{k};\overline{x}_{1},\cdots ,%
\overline{x}_{k};\lambda ,\Lambda _{L},\mu )=\left\langle \varphi
(x_{1})\cdots \varphi (x_{k})\right\rangle ,
\]
we have 
\begin{equation}
\left( \mu \frac{\partial }{\partial \mu }+\widehat{\beta }_{\lambda }\frac{%
\partial }{\partial \lambda }+\eta _{L}\Lambda _{L}\frac{\partial }{\partial
\Lambda _{L}}+k\gamma _{\varphi }\right) G_{k}(\widehat{x}_{1},\cdots ,%
\widehat{x}_{k};\overline{x}_{1},\cdots ,\overline{x}_{k};\lambda ,\Lambda
_{L},\mu )=0.  \label{CS}
\end{equation}
The equation can be immediately integrated to give 
\[
G_{k}(\widehat{x}_{1},\cdots ,\widehat{x}_{k};\overline{x}_{1},\cdots ,%
\overline{x}_{k};\lambda ,\Lambda _{L},\xi \mu )=z^{-k}(t)G_{k}(\widehat{x}%
_{1},\cdots ,\widehat{x}_{k};\overline{x}_{1},\cdots ,\overline{x}%
_{k};\lambda (t),\Lambda _{L}(t),\mu ),
\]
where $t=\ln \xi $ and 
\[
z(t)=\exp \left( \int_{0}^{t}\gamma _{\varphi }(\lambda (t^{\prime }))%
\mathrm{d}t^{\prime }\right) ,\qquad \frac{\mathrm{d}\lambda (t)}{\mathrm{d}t%
}=-\widehat{\beta }_{\lambda }(\lambda (t)),\qquad \Lambda _{L}(t)=\Lambda
_{L}\exp \left( -\int_{0}^{t}\eta _{L}(\lambda (t^{\prime }))\mathrm{d}%
t^{\prime }\right) ,
\]
with $\lambda (0)=\lambda $. Now the renormalization-group flow specifies
how the correlation functions changes under a weighted overall rescaling.
Indeed, the weighted scale invariance (\ref{wd2})-(\ref{ag}) tells us that 
\[
G_{k}(\widehat{x}_{1},\cdots ,\widehat{x}_{k};\overline{x}_{1},\cdots ,%
\overline{x}_{k};\lambda ,\Lambda _{L},\xi \mu )=\xi ^{k(\text{\DH }%
-2)/2}G_{k}(\xi \widehat{x}_{1},\cdots ,\xi \widehat{x}_{k};\xi ^{1/n}%
\overline{x}_{1},\cdots ,\xi ^{1/n}\overline{x}_{k};\lambda ,\Lambda
_{L},\mu ).
\]

A one-loop calculation for the models (\ref{22n}) gives 
\[
\widehat{\beta }_{\lambda }=-\varepsilon \lambda +\frac{3\lambda ^{2}}{(4\pi
)^{n+1}n!}+\mathcal{O}(\lambda ^{3}),\qquad \gamma _{\varphi }=\mathcal{O}%
(\lambda ^{2}),\qquad \eta _{L}=\mathcal{O}(\lambda ^{2}), 
\]
so these models are IR free. Only the beta function has a nonvanishing
one-loop contribution. Indeed, using the dimensional-regularization
technique tadpoles vanish in homogeneous models, so $\gamma _{\varphi }$ and 
$\eta _{L}$ start from two loops.

Let us now consider the model (\ref{4d}). The bare lagrangian reads 
\[
\mathcal{L}_{(2,2)\mathrm{B}}=\frac{1}{2}(\widehat{\partial }\varphi _{%
\mathrm{B}})^{2}+\frac{1}{2\Lambda _{L\mathrm{B}}^{2}}\left( \overline{%
\triangle }\varphi _{\mathrm{B}}\right) ^{2}+\frac{\lambda _{4\mathrm{B}}}{%
4!\Lambda _{L\mathrm{B}}^{2-\varepsilon _{2}/2}}\varphi _{\mathrm{B}}^{2}(%
\overline{\partial }\varphi _{\mathrm{B}})^{2}+\frac{\lambda _{6\mathrm{B}}}{%
6!\Lambda _{L\mathrm{B}}^{2-\varepsilon _{2}}}\varphi _{\mathrm{B}}^{6}, 
\]
where 
\[
\varphi _{\mathrm{B}}=Z_{\varphi }^{1/2}\varphi ,\quad \Lambda _{L\mathrm{B}%
}=Z_{\Lambda }\Lambda _{L},\quad \lambda _{4\mathrm{B}}=\mu ^{\varepsilon
}\left( \lambda _{4}+\Delta _{4}\right) ,\quad \lambda _{6\mathrm{B}}=\mu
^{2\varepsilon }\left( \lambda _{6}+\Delta _{6}\right) ,\quad \varepsilon
\equiv \varepsilon _{1}+\frac{\varepsilon _{2}}{2}. 
\]
The theory is invariant under the scale transformation (\ref{wd2}) with $n=2$%
. At one-loop we find $Z_{\varphi }=1,$ $Z_{\Lambda }=1$ and 
\[
\Delta _{4}=\frac{5\lambda _{4}^{2}}{2(12\pi )^{2}\varepsilon },\qquad
\Delta _{6}=\frac{5\lambda _{4}\lambda _{6}}{(8\pi )^{2}\varepsilon }-\frac{%
5\lambda _{4}^{3}}{(48\pi )^{2}\varepsilon }, 
\]
so the beta functions read 
\[
\widehat{\beta }_{4}=-\varepsilon \lambda _{4}+\frac{5\lambda _{4}^{2}}{%
2(12\pi )^{2}},\qquad \widehat{\beta }_{6}=-2\varepsilon \lambda _{6}+\frac{%
5\lambda _{4}\lambda _{6}}{(8\pi )^{2}}-\frac{5\lambda _{4}^{3}}{(48\pi )^{2}%
}. 
\]
The asymptotic solutions of the RG flow equations are 
\[
\lambda _{4}\sim \frac{2(12\pi )^{2}}{5t},\qquad \lambda _{6}\sim \frac{1}{20%
}\lambda _{4}^{2}, 
\]
where $t=\ln |x|\mu $ and $|x|$ is a typical weighted scale of the process.
Since $\lambda _{4}$ and $\lambda _{6}$ must be nonnegative, the theory is
IR\ free.

\section{Weighted trace anomaly}

\setcounter{equation}{0}

The weighted scale invariance (\ref{wd}) of the homogeneous models can be
anomalous due to the radiative corrections. In this section we calculate the
weighted trace anomaly, following \cite{hathrell}. For definiteness, we work
with the model (\ref{six}), but the discussion generalizes immediately to
the other models.

\paragraph{Weighted dilatation.}

In the case of the model (\ref{six}), write the lagrangian as $\mathcal{L}%
(\varphi ,\widehat{\partial }_{\mu }\varphi ,\overline{\triangle }\varphi )$%
. The infinitesimal version of the transformation (\ref{wd}) reads 
\[
\delta \varphi =\Omega \left( 1+\widehat{x}\cdot \widehat{\partial }+\frac{1%
}{2}\overline{x}\cdot \overline{\partial }\right) \varphi \equiv \Omega %
\hbox{\it \v{D}}\varphi , 
\]
with $\Omega \ll 1$. The conserved Noether current $J^{\mu }=(\widehat{J}%
^{\mu },\overline{J}^{\mu })$ is given by 
\[
\widehat{J}^{\mu }=-\widehat{x}^{\mu }\mathcal{L}+\frac{\partial \mathcal{L}%
}{\partial (\widehat{\partial }_{\mu }\varphi )}\hbox{\it \v{D}}\varphi
,\qquad \overline{J}^{\mu }=-\frac{1}{2}\overline{x}^{\mu }\mathcal{L}+\frac{%
\partial \mathcal{L}}{\partial (\overline{\triangle }\varphi )}%
\overleftrightarrow{\overline{\partial }^{\mu }}\hbox{\it \v{D}}\varphi . 
\]
We continue the spacetime dimensions to complex values as explained in
section 1. The continued transformation $\delta \varphi ^{\prime }$ and the
continued current $J^{\prime \hspace{0.01in}\mu }$ are obtained replacing 
\textit{\v{D}}$\varphi $ in $\delta \varphi $ and $J^{\mu }$ with 
\begin{equation}
\hbox{\it \v{D}}^{\prime }\varphi =\left( \frac{\hbox{\it \DH }}{2}-1+%
\widehat{x}\cdot \widehat{\partial }+\frac{1}{2}\overline{x}\cdot \overline{%
\partial }\right) \varphi  \label{sym}
\end{equation}
(see (\ref{wd2})), where \textit{\DH }$=4-\varepsilon $. At the bare level,
the anomaly of (\ref{sym}) is expressed by the divergence of $J^{\prime 
\hspace{0.01in}\mu }$. We find 
\begin{equation}
\partial _{\mu }J^{\prime \mathrm{\hspace{0.01in}}\mu }=-\varepsilon \frac{%
\lambda _{\mathrm{B}}\varphi _{\mathrm{B}}^{4}}{4!\Lambda _{\mathrm{B}L}^{2}}%
.  \label{diva}
\end{equation}

\paragraph{Improved energy-momentum tensor and its weighted trace.}

The anomaly of the weighted dilatation is encoded also in the
energy-momentum tensor, precisely in its ``weighted trace''. Let us start
from the energy-momentum tensor given by the Noether method. For the model (%
\ref{six}) we have 
\begin{equation}
T_{\mu \nu }=\frac{\partial \mathcal{L}}{\partial (\widehat{\partial }_{\mu
}\varphi )}\partial _{\nu }\varphi +\frac{\partial \mathcal{L}}{\partial (%
\overline{\triangle }\varphi )}\overleftrightarrow{\overline{\partial }_{\mu
}}\partial _{\nu }\varphi -\delta _{\mu \nu }\mathcal{L}.  \label{noe}
\end{equation}
This tensor is not symmetric, but conserved: it is easy to check that $%
\partial _{\mu }T_{\mu \nu }=0$, using the field equations. Next, define the
improved energy-momentum tensor 
\begin{eqnarray}
\widetilde{T}_{\mu \nu } &=&\widehat{\partial }_{\mu }\varphi \partial _{\nu
}\varphi -\frac{1}{\Lambda _{L}^{2}}\partial _{\nu }\varphi 
\overleftrightarrow{\ \overline{\partial }_{\mu }}\overline{\triangle }%
\varphi -\delta _{\mu \nu }\mathcal{L}-\frac{\hbox{\it \DH }-2}{4(\widehat{D}%
-1)}\widehat{\pi }_{\mu \nu }\varphi ^{2}+\frac{3\hbox{\it \DH }-2\overline{D%
}\hbox{\it \DH }+3\overline{D}-5}{(\overline{D}-1)\Lambda _{L}^{2}}\overline{%
\pi }_{\mu \nu }\left( \varphi \overline{\triangle }\varphi \right) 
\nonumber \\
&&+\frac{3-2\hbox{\it \DH }}{2(\overline{D}-1)\Lambda _{L}^{2}}\overline{\pi 
}_{\mu \nu }\left( \overline{\partial }_{\alpha }\varphi \right) ^{2}+\frac{%
3-2\hbox{\it \DH }}{\Lambda _{L}^{2}}\overline{\pi }_{\mu \alpha }\left(
\varphi \overline{\pi }_{\alpha \nu }\varphi \right) .  \label{stress}
\end{eqnarray}
where $\widehat{\pi }_{\mu \nu }=\widehat{\partial }_{\mu }\widehat{\partial 
}_{\nu }-\widehat{\delta }_{\mu \nu }\widehat{\partial }^{2}$ and $\overline{%
\pi }_{\mu \nu }=\overline{\partial }_{\mu }\overline{\partial }_{\nu }-%
\overline{\delta }_{\mu \nu }\overline{\partial }^{2}$. The first three
terms of (\ref{stress}) correspond to the Noether tensor (\ref{noe}), while
the rest collects the improvement terms, identically conserved. Define the
weighted trace 
\[
\Theta \equiv \widetilde{T}_{\widehat{\mu }\widehat{\mu }}+\frac{1}{n}%
\widetilde{T}_{\overline{\mu }\overline{\mu }}. 
\]
Using the field equations, it is easy to show that $\widetilde{T}_{\mu \nu }$
is conserved and that its weighted trace $\Theta $ vanishes in the physical
spacetime dimension $d=\widehat{d}+\overline{d}$. Moreover, $\widetilde{T}%
_{\mu \nu }$ is conserved also in the continued spacetime dimension. The
coefficients of the improvement terms are chosen so that in the free-field
limit $\Theta $ vanishes also in the continued dimension $D=\widehat{D}+%
\overline{D}$. Finally, it is straightforward to check that the weighted trace $%
\Theta $ coincides with the divergence (\ref{diva}) of the current $%
J^{\prime \hspace{0.01in}\mu }$.

\paragraph{Anomaly.}

We need to write $\Theta $ in terms of renormalized operators. When we
differentiate a renormalized correlation function with respect to $\lambda $
or $\Lambda _{L}$ we obtain a renormalized correlation function containing
additional insertions of $-\partial S/\partial \lambda $ or $-\partial
S/\partial \Lambda _{L}$, respectively. Thus, $-\partial S/\partial \lambda $
and $-\partial S/\partial \Lambda _{L}$ are renormalized operators.
Following a standard procedure \cite{hathrell} we can find which operators $%
\mathcal{O}$ they are the renormalized versions of. In the minimal
subtraction scheme, it is sufficient to express the renormalized operators
as bare operators $\mathcal{O}_{\mathrm{B}}$ plus poles. Schematically, 
\[
\mathrm{finite}=\mathcal{O}_{\mathrm{B}}+\mathrm{poles\qquad }\Rightarrow
\qquad \mathrm{finite}=[\mathcal{O}]. 
\]
where $[\mathcal{O}]$ denotes the renormalized version of the operator $%
\mathcal{O}$. We find 
\begin{eqnarray*}
\frac{\partial S}{\partial \lambda } &=&\mathrm{finite}=\frac{1}{\widehat{%
\beta }_{\lambda }}\left( \gamma _{\varphi }[E_{\varphi }]-\eta _{L}\Lambda
_{L}\frac{\partial S}{\partial \Lambda _{L}}-\varepsilon \frac{\lambda _{%
\mathrm{B}}}{4!\Lambda _{\mathrm{B}L}^{2}}\int \varphi _{\mathrm{B}%
}^{4}\right) =\frac{\mu ^{\varepsilon }}{4!\Lambda _{L}^{2}}\int [\varphi
^{4}], \\
-\frac{1}{2}\Lambda _{L}\frac{\partial S}{\partial \Lambda _{L}} &=&\mathrm{%
finite}=\frac{1}{2\Lambda _{\mathrm{B}L}^{2}}\int (\overline{\triangle }%
\varphi _{\mathrm{B}})^{2}+\frac{\lambda _{\mathrm{B}}}{4!\Lambda _{\mathrm{B%
}L}^{2}}\int \varphi _{\mathrm{B}}^{4}=\frac{1}{2\Lambda _{L}^{2}}\int [(%
\overline{\triangle }\varphi )^{2}]+\frac{\lambda \mu ^{\varepsilon }}{%
4!\Lambda _{L}^{2}}\int [\varphi ^{4}],
\end{eqnarray*}
where $[E_{\varphi }]=\int \varphi (\delta S/\delta \varphi )$ is the
operator that counts the number of $\varphi $-insertions. Thus, 
\[
\int \Theta =-\int \varepsilon \frac{\lambda _{\mathrm{B}}\varphi _{\mathrm{B%
}}^{4}}{4!\Lambda _{\mathrm{B}L}^{2}}=\frac{(\widehat{\beta }_{\lambda
}-2\lambda \eta _{L})\mu ^{\varepsilon }}{4!\Lambda _{L}^{2}}\int [\varphi
^{4}]-\frac{\eta _{L}}{\Lambda _{L}^{2}}\int [(\overline{\triangle }\varphi
)^{2}]-\gamma _{\varphi }[E_{\varphi }]. 
\]
The result agrees with the Callan-Symanzik equation (\ref{CS}), which can be
expressed as 
\[
\left\langle \int \Theta \ \varphi (x_{1})\cdots \varphi
(x_{k})\right\rangle =\mu \frac{\partial }{\partial \mu }\left\langle
\varphi (x_{1})\cdots \varphi (x_{k})\right\rangle . 
\]
Indeed, 
\[
\int \Theta =-\mu \frac{\partial S}{\partial \mu }=\widehat{\beta }_{\lambda
}\frac{\partial S}{\partial \lambda }+\eta _{L}\Lambda _{L}\frac{\partial S}{%
\partial \Lambda _{L}}-\gamma _{\varphi }[E_{\varphi }]. 
\]

\section{Nonrelativistic theories}

\setcounter{equation}{0}

Nonrelativistic theories can be studied along the same lines. The action
contains only a single time derivative $\widehat{\partial }$, 
\[
\mathcal{L}=\overline{\varphi }\left( \widehat{\partial }+\frac{\overline{%
\triangle }}{2m}+\xi \frac{\overline{\triangle }^{2}}{m^{2}}+\cdots \right)
\varphi +\zeta \overline{\varphi }^{2}\overline{\triangle }\varphi
^{2}+\cdots +\lambda (\overline{\varphi }\varphi )^{2}+\cdots 
\]
so the theory is more divergent. The dimensional-regularization is not easy
to use, since there is no simple way to continue the single-derivative term $%
\overline{\varphi }\widehat{\partial }\varphi $ to complex dimensions. Thus
we assume an ordinary cut-off regularization.

The propagator is defined by the term $\overline{\varphi }\widehat{\partial }%
\varphi $ plus the lagrangian quadratic term with the highest number of $%
\overline{\partial }$-derivatives, say $2n$, 
\[
\mathcal{L}_{\mathrm{free}}=\overline{\varphi }\left( \widehat{\partial }+%
\frac{\overline{\partial }^{2n}}{\Lambda _{L}^{2n-1}}\right) \varphi . 
\]
For the purposes of renormalization, the other quadratic terms, if present,
can be treated perturbatively, as explained in section 2. Thus the
nonrelativistic propagator is the inverse of a homogeneous weighted
polynomial of degree 1 and weight $1/n$. The integral measure has weighted
degree \textit{\dj }$=1+(d-1)/(2n)$. A Feynman diagram $G$ with $E$ total
external legs, $I$ propagators and $v_{N}^{(\alpha )}$ $N$-leg vertices of
weighted degrees $\delta _{N}^{(\alpha )}$ is a weighted function of degree 
\[
\omega (G)=L\hbox{\it \dj }-I+\sum_{(N,\alpha )}\delta _{N}^{(\alpha
)}v_{N}^{(\alpha )}. 
\]
Formulas (\ref{standard}) still hold. We have 
\[
\omega (G)=\hbox{\it \dj }-\frac{E}{2}\left( \hbox{\it \dj }-1\right)
+\sum_{(N,\alpha )}\left[ \delta _{N}^{(\alpha )}+\left( \frac{N}{2}%
-1\right) \hbox{\it \dj }-\frac{N}{2}\right] v_{N}^{(\alpha )}. 
\]
Renormalizable theories are those that contain the vertices with 
\begin{equation}
\delta _{N}^{(\alpha )}\leq \frac{N}{2}-\left( \frac{N}{2}-1\right) 
\hbox
{\it \dj }.  \label{con}
\end{equation}
Strictly renormalizable theories are those that have 
\[
\delta _{N}^{(\alpha )}=\frac{N}{2}-\left( \frac{N}{2}-1\right) 
\hbox {\it
\dj }. 
\]
Polynomiality requires now 
\[
\hbox{\it \dj }>1, 
\]
which ensures also that $\omega (G)$ decreases when the number of external
legs increases. The maximal number of legs is 
\begin{equation}
N_{\mathrm{max}}=\left[ \frac{2\hbox{\it \dj }}{\hbox{\it \dj }-1}\right] .
\label{nmax2}
\end{equation}
It is straightforward to check that $E=N$ implies

\[
\omega (G)\leq \hbox{\it \dj }-\frac{N}{2}\left( \hbox{\it \dj }-1\right) , 
\]
so by (\ref{con}) the type of vertex that subtracts the divergence of $G$ is
already present in the lagrangian, which proves renormalizability. No terms
with more than one time derivative are turned on by renormalization.

Let us now see some examples of homogeneous models, beginning from the $%
\varphi ^{4}$-theories. Setting $N_{\mathrm{max}}=4$ in (\ref{nmax2}) we get 
\[
\frac{5}{3}<\hbox{\it \dj }\leq 2. 
\]
For \textit{\dj }$=2$ we have $d=2n+1$ and the family of odd-dimensional
theories 
\begin{equation}
\mathcal{L}_{(1,2n)}=\overline{\varphi }i\widehat{\partial }\varphi +\frac{1%
}{\Lambda _{L}^{2n-1}}\overline{\varphi }\overline{\partial }^{2n}\varphi +%
\frac{\lambda }{4\Lambda _{L}^{2n-1}}(\overline{\varphi }\varphi )^{2}.
\label{nonrel}
\end{equation}
Setting $N_{\mathrm{max}}=6$ we have $7/5<$\textit{\dj }$\leq 3/2$. For 
\textit{\dj }$=3/2$ we have $d=n+1$. If $n$ is odd we have the family 
\[
\mathcal{L}_{(1,n)}=\overline{\varphi }i\widehat{\partial }\varphi +\frac{1}{%
\Lambda _{L}^{2n-1}}\overline{\varphi }\overline{\partial }^{2n}\varphi +%
\frac{\lambda _{6}}{36\Lambda _{L}^{2n-1}}(\overline{\varphi }\varphi )^{3}. 
\]
In particular, we see that there exist four-dimensional ($n=3$)
nonrelativistic renormalizable $\varphi ^{6}$-theories. If $n$ is even we
must include additional vertices, 
\[
\mathcal{L}_{(1,n)}=\overline{\varphi }i\widehat{\partial }\varphi +\frac{1}{%
\Lambda _{L}^{2n-1}}\overline{\varphi }\overline{\partial }^{2n}\varphi
+\sum_{\beta }\frac{\lambda _{\beta }}{4\Lambda _{L}^{2n-1}}[\overline{%
\partial }^{n}\overline{\varphi }^{2}\varphi ^{2}]_{\beta }+\frac{\lambda
_{6}}{36\Lambda _{L}^{2n-1}}(\overline{\varphi }\varphi )^{3}. 
\]

\section{Conclusions}

\setcounter{equation}{0}

In this paper we have classified the unitary Lorentz violating
renormalizable quantum field theories that can be obtained improving the UV
behavior of propagators with the help of higher space derivatives. The
removal of divergences is governed by a weighted power-counting criterion.
If the lagrangian has an appropriate form, time derivatives are
``protected'', in the sense that no higher time derivatives are turned on by
renormalization. The so-defined theories are unitarity, but have modified
dispersion relations. We have studied their main properties, including the
renormalization group flow and the weighted trace anomaly.

Natural extensions of this work are those that aim to include gauge fields
and gravity. Possible applications range from high-energy physics, effective
field theory, nuclear physics and the theory of critical phenomena. In the
high-energy physics domain, it would be interesting to explore the
work-hypothesis that Lorentz invariance is violated at very high energies,
to define the ultraviolet limit of quantum gravity, or study new types of
Lorentz invariant extensions of the Standard Model. It would also be
interesting to embed the weighted scale invariance into a ``weighted
conformal group'', generalizing the Galilean conformal group that
characterizes a class of nonrelativistic theories \cite{wise2}.

\vskip 20truept \noindent {\Large \textbf{Appendix A: extra subdivergences}}

\vskip 10truept

\renewcommand{\theequation}{A.\arabic{equation}} \setcounter{equation}{0}

In this appendix we give more details on the extra subdivergences mentioned
in section 4. By construction, every row of table (\ref{table}) is free of
``ordinary'' subdivergences, namely those originated by the subdiagrams $%
\gamma $. Every column is free of overall divergences. Extra subdivergences
are those that occur when a subdiagram $\gamma ^{\prime }$ is convergent in $%
Q_{G}$, but becomes divergent in one of the $\overline{Q}_{\gamma }$'s,
after replacing $\gamma $ with its counterterms. Here we prove that the
extra subdivergences are automatically subtracted column-wise.

It is useful to have an explicit example in mind, such as the two-loop
diagram depicted in fig. 1, in the four dimensional $\varphi ^{4}$-theory.
The diagram is the $p$-$q$ integral of 
\[
Q_{G}=\frac{1}{(p^{2}+m^{2})\left[ (p-k)^{2}+m^{2}\right] }\frac{1}{%
(q^{2}+m^{2})\left[ (q+p+k^{\prime })^{2}+m^{2}\right] }.
\]
The $p$-integral is convergent, the $q$-integral is not. The $q$%
-subdivergence is subtracted by 
\begin{equation}
-\overline{Q}_{\gamma }=-\frac{1}{(p^{2}+m^{2})\left[ (p-k)^{2}+m^{2}\right] 
}\frac{1}{(q^{2}+m^{2})^{2}}.  \label{col}
\end{equation}
In this expression, however, the $p$-integral is divergent. This divergence
is what we call an extra subdivergence. The table reads 
\begin{equation}
\begin{tabular}{ccc}
$Q_{G}$ & \phantom{CCC} & $-\overline{Q}_{\gamma }$ \\ 
$-\frac{1}{(p^{2}+m^{2})^{2}}\frac{1}{(q^{2}+m^{2})\left[
(q+p)^{2}+m^{2}\right] }$ & \phantom{CCC} & $+\frac{1}{(p^{2}+m^{2})^{2}}%
\frac{1}{(q^{2}+m^{2})^{2}}$%
\end{tabular}
\label{tabla}
\end{equation}

\begin{figure}[tbp]
\centerline{\includegraphics[width=1.5in,height=1in]{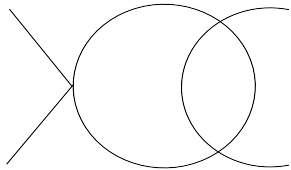}}
\caption{Simple example of diagram that generates an extra subdivergence}
\end{figure}

In the general case, assume that the subdiagram $\gamma $ contains $l$ loops
and that $\overline{Q}_{\gamma }$ contains some extra subdivergences. The
extra subdivergences can be overall or not. We call them overall if they
arise letting all of the remaining $L-l$ loop momenta tend to infinity. They
are not overall if they arise letting only a subset of the remaining $L-l$
loop momenta tend to infinity. Proceeding inductively, we can assume that
the non-overall extra subdivergences have already been subtracted away.
Thus, we need to consider only the overall extra subdivergences. It is not
difficult to see that they are subtracted column-wise in (\ref{table}).
Indeed, as in (\ref{col}), the integrands that generate extra overall
subdivergences factorize (or split into a sum of terms each of which
factorizes): one factor is responsible for the extra subdivergence (see the
first factor of $-\overline{Q}_{\gamma }$ in (\ref{col})), while the other
factor is the $\gamma $-counterterm (see the second factor of $-\overline{Q}%
_{\gamma }$ in (\ref{col})). The second factor is the same throughout the
column. Thus, the column subtracts away the overall divergence of the first
factor, which is precisely the extra subdivergence. Recapitulating, the rows
are free of ordinary subdivergences and the columns are free of extra
subdivergences and overall divergences. Thus the table (\ref{table}) is
convergent. In the example (\ref{tabla}), it is clear that the column of $-%
\overline{Q}_{\gamma }$ is $p$-convergent.

\vskip 20truept \noindent {\Large \textbf{Appendix B: Euclidean propagators}}

\vskip 10truept

\renewcommand{\theequation}{B.\arabic{equation}} \setcounter{equation}{0}

Let us examine some propagators 
\[
\frac{1}{\widehat{p}^{2}+\frac{(\overline{p}^{2})^{n}}{\Lambda _{L}^{2n-2}}} 
\]
in coordinate space. The Euclidean (2,2)-propagator in four dimensions with $%
n=2$ reads

\[
G_{(2,2)}(\widehat{x},\overline{x},\Lambda _{L})=\frac{\Lambda _{L}}{16|%
\widehat{x}|}\left[ I_{0}(\Lambda _{L}\overline{x}^{2}/4|\widehat{x}%
|)-SL_{0}(\Lambda _{L}\overline{x}^{2}/4|\widehat{x}|)\right] ,
\]
where $I$ denotes the modified Bessel function of the first kind, while $SL$
denotes the modified Struve function. For $|\widehat{x}|\gg \Lambda _{L}%
\overline{x}^{2}$ and $|\widehat{x}|\ll \Lambda _{L}\overline{x}^{2}$ we
have 
\[
G_{(2,2)}\sim \frac{\Lambda _{L}}{16|\widehat{x}|}\qquad \mathrm{and}\qquad
G_{(2,2)}\sim \frac{1}{2\pi \overline{x}^{2}},
\]
respectively.

Instead, the Euclidean (1,3)-propagator with $n=2$ reads 
\[
G_{(1,3)}(\widehat{x},\overline{x},\Lambda _{L})=\frac{\Lambda _{L}}{8\pi |%
\overline{x}|}\mathrm{Erf}\left( \sqrt{\frac{\Lambda _{L}\overline{x}^{2}}{4|%
\widehat{x}|}}\right) . 
\]
In the two limits considered above we have the behaviors 
\[
G_{(1,3)}\sim \frac{\Lambda _{L}^{3/2}}{8\pi ^{3/2}|\widehat{x}|^{1/2}}%
\qquad \mathrm{and}\qquad G_{(1,3)}\sim \frac{\Lambda _{L}}{8\pi |\overline{x%
}|}, 
\]
respectively.

\end{document}